\documentclass[twocolumn,aps,prb,amsmath,amssymb,superscriptaddress,showpacs]{revtex4}

\usepackage{graphicx} % Include figure files

\begin{document}

\title{Two-dimensional vortex behavior in highly underdoped YBa$_2$Cu$_3$O$_{6+x}$ observed by scanning Hall probe microscopy}

\author{J. W. Guikema}
\altaffiliation[Address as of March 2008: ]{Department of Physics \& Astronomy, Johns Hopkins University, Baltimore, MD 21218}
\author{Hendrik Bluhm}
\affiliation{Departments of Physics and of Applied Physics,
Stanford University, Stanford, CA 94305}
\author{D. A. Bonn}
\author{Ruixing Liang}
\author{W. N. Hardy}
\affiliation{Department of Physics and Astronomy, University of
British Columbia, Vancouver, BC, Canada V6T 1Z1}
\author{K. A. Moler}
\email{kmoler@stanford.edu} \affiliation{Departments of Physics
and of Applied Physics, Stanford University, Stanford, CA 94305}

\date{February 10, 2008}

\begin{abstract}
We report scanning Hall probe microscopy of highly underdoped
superconducting \mbox{YBa$_2$Cu$_3$O$_{6+x}$} with $T_c$ ranging
from 5 to 15~K which showed distinct flux bundles with less than
one superconducting flux quantum ($\Phi_0$) through the sample
surface.  The sub-$\Phi_0$ features occurred more frequently for
lower $T_c$, were more mobile than conventional vortices, and
occurred more readily when the sample was cooled with an in-plane
field component.  We show that these features are consistent with
kinked stacks of pancake vortices.
\end{abstract}

% 74.25.Ha    Magnetic properties
% 74.25.Qt    Vortex lattices, flux pinning, flux creep
% 74.72.Bk    Y-based cuprates

\maketitle

% Define shortcut commands to be used below
\newcommand{\micron}{\ensuremath{\mu}\mbox{m}}
\newcommand{\ybco}[1]{\mbox{YBa$_2$Cu$_3$O$_{#1}$}}
\newcommand{\degC}{$^{\circ}\mathrm{C}$}

\section{Introduction}

Cuprate superconductivity occurs mainly in the $ab$ direction on
the $\mathrm{CuO}_2$ planes.  This quasi-two-dimensional
nature manifests itself in the anisotropy between the $c$-axis
penetration depth ($\lambda_c$) and the in-plane penetration depth
($\lambda_{ab}$). Clem\cite{ClemJR:Two-di} showed that in highly
anisotropic layered superconductors a $c$-axis vortex can be
viewed as a stack of magnetically coupled ``pancake'' vortices,
one in each layer. This formulation suggested the possibility for
novel vortex behavior and has become a major part of the
phenomenological understanding of cuprate superconductors.
However, direct observations of separated pancakes or pancake
stacks have been rare.\cite{GrigorenkoAN:Visicv,Beleggia}

In this paper, we present scanned probe microscopy of magnetic flux
in highly underdoped \ybco{6+x} (YBCO) single crystals.  We observed
nearly isolated flux features with less than one flux quantum
($\Phi_0 = hc/2e = 20.7$~G\,\micron$^2$) through the sample
surface, which we call ``partial vortices''.  A model of separated
pancake vortex stacks, similar to a kinked structure suggested by
Benkraouda and Clem,\cite{BenkraoudaM:Instvl} but with a more
important role for pinning, agrees well with our observations.

Non-quantized flux in superconductors has been observed experimentally, arising for different reasons.  Geim et al.\cite{Geim_Nature_2000}\ observed non-quantized flux penetration in mesoscopic thin film samples of aluminum due to two effects: the proximity of the vortices to the sample edge, and a surface barrier to flux penetration.  Sub-$\Phi_0$ flux has been imaged in YBCO thin films along grain boundaries separating regions of the crystal rotated 45$^{\circ}$ about the $c$-axis due to the $d$-wave symmetry of the pairing state in combination with facets along the grain boundary.\cite{KirtleyJR:Dismfh,MannhartJ:Genmfb} For our measurements discussed here, the vortices were far from the edge and there were no rotations of the crystal axes aside from 90$^{\circ}$ twinning, so neither of these mechanisms applies.  Our observations of seemingly isolated fractional fluxes in a bulk material far from any boundaries requires a different explanation.

This paper is organized as follows. Section \ref{sec:samples}
describes the growth and preparation of the high-quality highly underdoped YBCO crystals and introduces the scanning Hall probe microscope.  Section \ref{sec:fluximages} discusses Hall probe
observations of partial vortices, and in Sec.\ \ref{sec:themodel}
we model these partial vortices as kinked stacks of two-dimensional (2D) pancake
vortices. Section \ref{sec:furtherdetails} presents other
experimentally observed properties, which are all consistent with
the kinked vortex picture.  Finally, in Sec.\ \ref{sec:discussion}
we discuss the partial vortices in light of energy costs and
pinning. The Appendix of this paper discusses the in-plane
penetration depth extracted from fits to vortices in the YBCO\@.

\section{Samples and methods}\label{sec:samples}

The \ybco{6+x} crystals, with $x$ from 0.34 to 0.375, are grown
with a self-flux method in BaZrO$_3$ crucibles detailed
elsewhere.\cite{LiangRuixing:GrohqY,LiangR:PrechY}  After growth,
the desired oxygen content is set during a 900--930\degC{} anneal
in flowing oxygen, then oxygen inhomogeneities are removed during
a 1--2 week 570\degC{} anneal in a small tube with YBCO ceramic at
the same oxygen content.  Initially, after quenching to 0\degC,
the crystals are non-superconducting, but annealing at room
temperature allows the oxygen atoms to order into Ortho-II (every
other chain empty) chain fragments whose increasing length
provides the carrier doping in the CuO$_2$
planes.\cite{LiangR:PrechY} The superconducting transition temperature ($T_c$) increases with room temperature
annealing until saturation is reached after several weeks, giving
final $T_c$ values of 5--20~K with bulk susceptibility transition
widths (10\%--90\%) of less than 2~K\@.\cite{LiangR:PrechY} This
early generation of underdoped samples sometimes had a small
($<$2\%) volume fraction of the 50--60~K $T_c$ phase of Ortho-II
YBCO, as observed in magnetization
measurements.\cite{Liang:PrivComm} During the room temperature
annealing, a single crystal can be observed at a range of $T_c$
values.  The platelet shaped crystals are about
$1~\mathrm{mm}\times 1~\mathrm{mm}$ wide, with their surface
parallel to the $ab$-plane, and are typically 10--100~\micron{}
thick. As grown, the crystals have twinning boundaries, but they
can be detwinned under uniaxial pressure at elevated temperatures.

Our most detailed observations of sub-$\Phi_0$ partial vortices
were made with a scanning Hall probe
microscope,\cite{GuikemaJW:THESIS,ChangAM:ScaHpm,DavidovicD:Cordam,OralA:ScaHpm}
described in Ref.\ \onlinecite{GuikemaJW:THESIS}. Scanning Hall
probe microscopy of single vortices is an established
technique\cite{ChangAM:ScaHpm,DavidovicD:Cordam,OralA:ScaHpm}
first demonstrated by A. M. Chang et al.\cite{ChangAM:ScaHpm} Our
Hall probe was made from GaAs/AlGaAs two-dimensional electron gas
and had lithographic size $0.5~\micron \times 0.5~\micron$. The
Hall probe measures the perpendicular magnetic field in the active
area (with a constant offset). The Hall cross was covered by a thin film of gold which was grounded during operation.  This gate prevented any stray electric charges on the sample surface from perturbing the Hall signal.  For positioning in the $z$ direction, the probe is mounted on the end of a thin aluminum diving board which forms a parallel plate capacitor with a copper pad underneath it.  As the tip of the probe approaches the sample, we monitor the capacitance and can ideally determine the location of the sample surface to within 10 nm.   This touchdown procedure is repeated at multiple locations within the scan area and then the probe is scanned in a plane just above the sample surface.  The minimum height of the Hall cross active
area above the sample surface is determined by the sample-probe
alignment and for these measurements was 0.4~\micron{} or larger due to geometric
constraints. The lateral scan range of the microscope at 4~K is
60~\micron{} and the sample can also be repositioned using $xy$ stick-slip course motion. For improved signal-to-noise ratio, we averaged multiple
images (having checked that the consecutive images did not show
changes). We sometimes used an $8~\micron \times 8~\micron$
scanning superconducting quantum interference device (SQUID) with
better flux sensitivity, but worse spatial resolution than the
Hall probe.  The cryostat was inside triple-layer mu-metal
magnetic shielding with a residual field of less than 25~mG\@.

\begin{figure}
\includegraphics{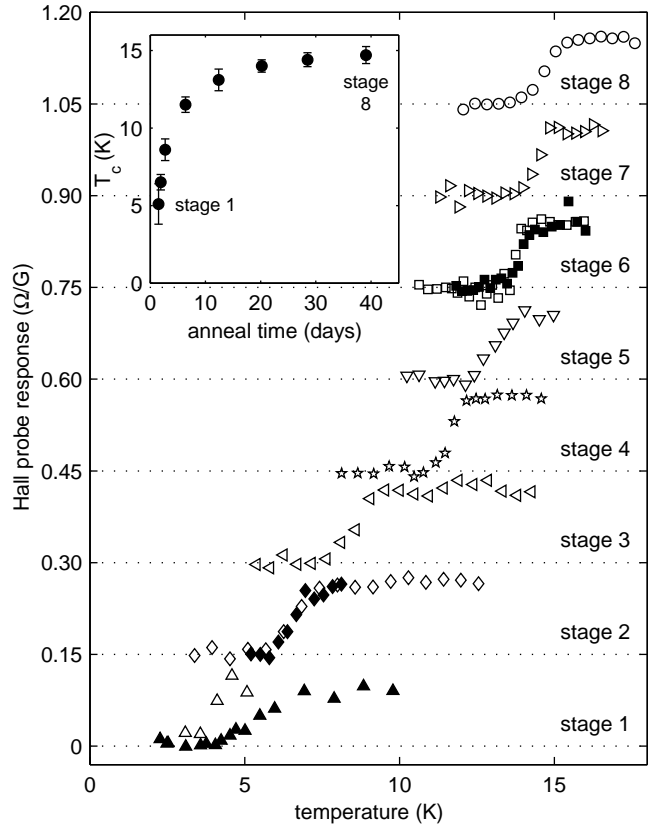}
\caption{\label{fig:TctransAnneal}Superconducting transitions in
the \ybco{6.375} crystal measured by in situ magnetic
susceptibility with the Hall probe in an applied 8.3~mHz field of
amplitude
0.25~Oe for stages 1--2 and 0.20~Oe for stages 3--8. Stages offset by $0.15~\Omega/$G\@. Filled
symbols are second measurements at different locations.  A
Hall probe response of $R_H=0.115\pm 0.015~\Omega/$G indicates no
measurable Meissner response, while $0~\Omega/$G indicates full
shielding. Inset: Midpoint transition temperature ($T_c$) versus
cumulative room temperature annealing time for the \ybco{6.375}
crystal. Vertical bars indicate the full transition width, limited
by our measurement resolution.}
\end{figure}

The main results presented in this paper are from an 8~\micron{}
thick twinned \ybco{6.375} crystal imaged at eight stages during
the room temperature oxygen ordering annealing.  The annealing took
place in the microscope in a helium atmosphere.  After 36 hours of
annealing, the crystal had $T_c\sim 5.1$~K and transition width
$\Delta T_c\sim 3$~K\@. Further annealing gave a range of $T_c$
values all having $\Delta T_c<1.5$~K\@. The maximum measured $T_c$
was 14.7~K (Fig.\ \ref{fig:TctransAnneal} inset).
%%%%%%%%%%%%%%%%%%%%%%%%%%%%%%%%%%
%%%%%%%%%%%%%%%%%%%%%%%%%%%%%%%%%%
$T_c$ values were obtained in situ in an 8.3~mHz applied field of
amplitude 0.20--0.25~Oe (Fig.\ \ref{fig:TctransAnneal}). The
transitions are described as midpoint $T_c$'s with full widths
limited by the $\sim$10\% resolution of the susceptibility
measurement.

We also imaged flux in nine other similarly prepared \ybco{6+x}
crystals with $x$ in the range 0.35--0.375 and $T_c$'s in the
range 7--17~K\@. Each of these crystals was studied at only one
$T_c$ value. Sub-$\Phi_0$ flux features were seen in the three
crystals with the lowest $T_c\sim 7$~K
values.\cite{GuikemaJW:THESIS} The higher $T_c$ samples with
$T_c\gtrsim 11$~K only showed flux consistent with conventional
$\Phi_0$ vortices.  We also imaged vortices in Ortho-II
($T_c\approx 60$~K) and near-optimally doped YBCO crystals and did
not see any evidence of sub-$\Phi_0$ partial vortices.

\section{Flux images}\label{sec:fluximages}

\begin{figure}
\includegraphics{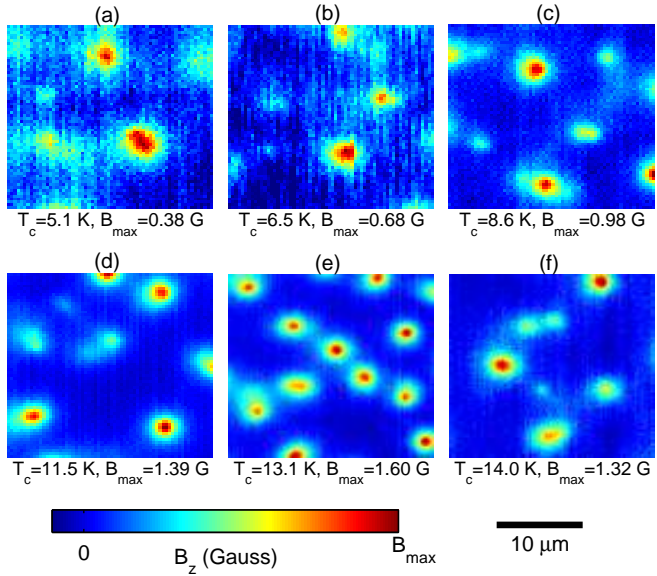}
\caption{\label{fig:HPpartials}(Color).  Hall probe images of
magnetic flux in the \ybco{6.375} crystal for increasing $T_c$
(stages 1--6). The probe's size was nominally 0.5~\micron.  The
circular features with largest $B_z$ are likely straight pancake
vortex stacks (``full vortices''), while the dimmer and sometimes
non-circular features are identified as segments of kinked pancake
vortex stacks (``partial vortices'').  The applied field $H_z$ while cooling through $T_c$ and the temperature $T$ at which each image was acquired were (a) 0.15 Oe and 2.0 K, (b) 0.21 Oe and 2.3 K, (c) 0.21 Oe and 2.4 K, (d) 0.21 Oe and 2.2 K, (e) 0.43 Oe and 4.3 K, and (f) 0.20 Oe and 4.1 K\@.  A constant
background has been subtracted from each image. For (f), low frequency telegraph noise in the Hall probe signal was subtracted from the raw images before averaging.}
\end{figure}

We saw over 100 sub-$\Phi_0$ flux features in the \ybco{6.375}
crystal while tuning $T_c$ from 5 to 15~K\@. We also observed more
than 300 apparently full $\Phi_0$ vortices, which were dominant
for $T_c>11.5$~K\@.  Typical Hall probe images are shown in Fig.\
\ref{fig:HPpartials} for a range of $T_c$.
%%%%%%%%%%%%%%%%%%%%%%%%%%%%%%%%%%
%%%%%%%%%%%%%%%%%%%%%%%%%%%%%%%%%%
These images were taken at low temperatures ($T<T_c/2$) after
field-cooling the crystal at about 3~K/min in a perpendicular
field.  Though not necessary, the images were taken after turning off the
applied field at low temperature.  Images taken before and after
turning off the field looked identical.  No flux features were
observed when we cooled the sample in zero field.  The images were not all taken at the same place on the crystal, because we occasionally used our course motion capability to move the sample in order to view nearby regions.

The images in Fig.\ \ref{fig:HPpartials} show flux features that
can be divided into two types.  We identify the brightest
features, which are close to circular, as conventional or ``full''
vortices. They carry total flux $\Phi_0$ through the crystal
surface, within experimental error. The full vortices increased in
peak $B_z$ and decreased in width as $T_c$ increased, likely due
to changes in the in-plane penetration depth (see Appendix). Other
features have a smaller peak $B_z$ and appear either circular,
elongated, or with tails. We call these features ``partial
vortices'' because they carry less than $\Phi_0$ of total flux
through the surface.  When the sample was cooled in a
perpendicular field, partial vortices accounted for more than half
of the observed flux features for $T_c\le 11.5$~K (stages 1--4),
but dropped to less than 10\% for $T_c > 14$~K (stages 7 and
8).\cite{GuikemaJW:THESIS}  We will show that the partial vortices
can be explained by non-axial arrangements of 2D pancake vortices.

\begin{figure}
\includegraphics{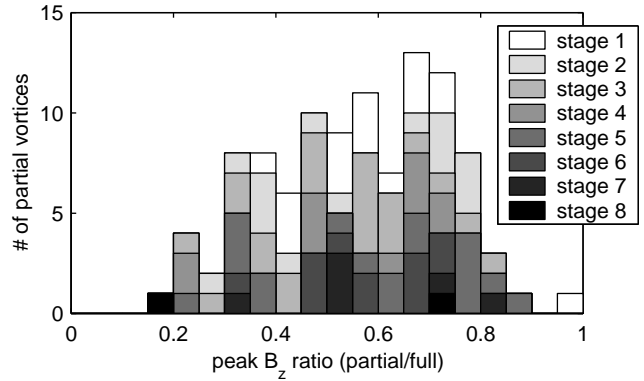}
\caption{\label{fig:PVhistogram}Histogram of partial vortex peak
field as a fraction of full vortex peak field in the \ybco{6.375}
crystal. Sample was cooled to low temperature in a perpendicular
field before each image was acquired. Total bar heights show data
from all anneal stages (all $T_c$ values).}
\end{figure}

The flux carried by the partial vortices was not restricted to
discrete fractions of $\Phi_0$, as shown by a tally of the ratios
of the peak $B_z$ of a partial vortex to the peak $B_z$ of a full
vortex in the same image (Fig.\ \ref{fig:PVhistogram}).
%%%%%%%%%%%%%%%%%%%%%%%%%%%%%%%%%%
%%%%%%%%%%%%%%%%%%%%%%%%%%%%%%%%%%
The peak $B_z$ ratio for a partial vortex does not translate
directly to the fraction of a $\Phi_0$ through the surface, but
nonetheless, Fig.\ \ref{fig:PVhistogram} indicates that the
partial vortices occurred for a range of magnitudes.  It is not feasible to directly tally the flux of each partial vortex without large errors because the field from nearby vortices in many cases interferes with the integration.

A tempting hypothesis is that a partial vortex consists of a straight stack of pancake vortices, one in each layer, with some unusual mechanism by which the total flux is permitted to be less than $\Phi_0$.  In this case, the field profile of a partial vortex above the sample surface could be calculated using the anisotropic London model (see Eq.\ (\ref{eqn:fullmodel}) in the Appendix) except with a flux smaller than $\Phi_0$.  If the in-plane penetration depth was assumed to be constant throughout the sample at each $T_c$, then the peak $B_z$ ratio given in Fig.\ \ref{fig:PVhistogram} would be equivalent to the fraction of a flux quantum carried by the partial vortex.  However, we do not believe this hypothesis is the best explanation of our data.

The main sources of error in
the peak $B_z$ ratios were noise in the data and error in
background determination, roughly 50~mG each. This gives an error
in the peak $B_z$ ratios of $\sim$5\% at the higher $T_c$ values,
and several times this for the lowest $T_c$. Counts may also be
missing at the ends of the histogram, since a partial vortex with
close to full peak would likely be mistaken as a full vortex and
those with very small peak field may have been lost in the noise.
We omitted some flux features from the histogram if they could not
be clearly identified as partial or full vortices.  We also
omitted whole images if they had high noise or did not contain a
full vortex, the latter being particularly an issue for the lowest
anneal stage where full vortices were rare. Only 21\% of the
images from stage 1 were included in the tally, while most images
from subsequent stages were included.

\section{Kinked pancake stacks}\label{sec:themodel}

In this section we discuss a model in which the partial vortices
result from kinked stacks of pancake vortices. We show that the
model quantitatively describes the most circular partial vortices,
and qualitatively describes the shapes and tails of the
non-circular ones.

\begin{figure*}
\includegraphics{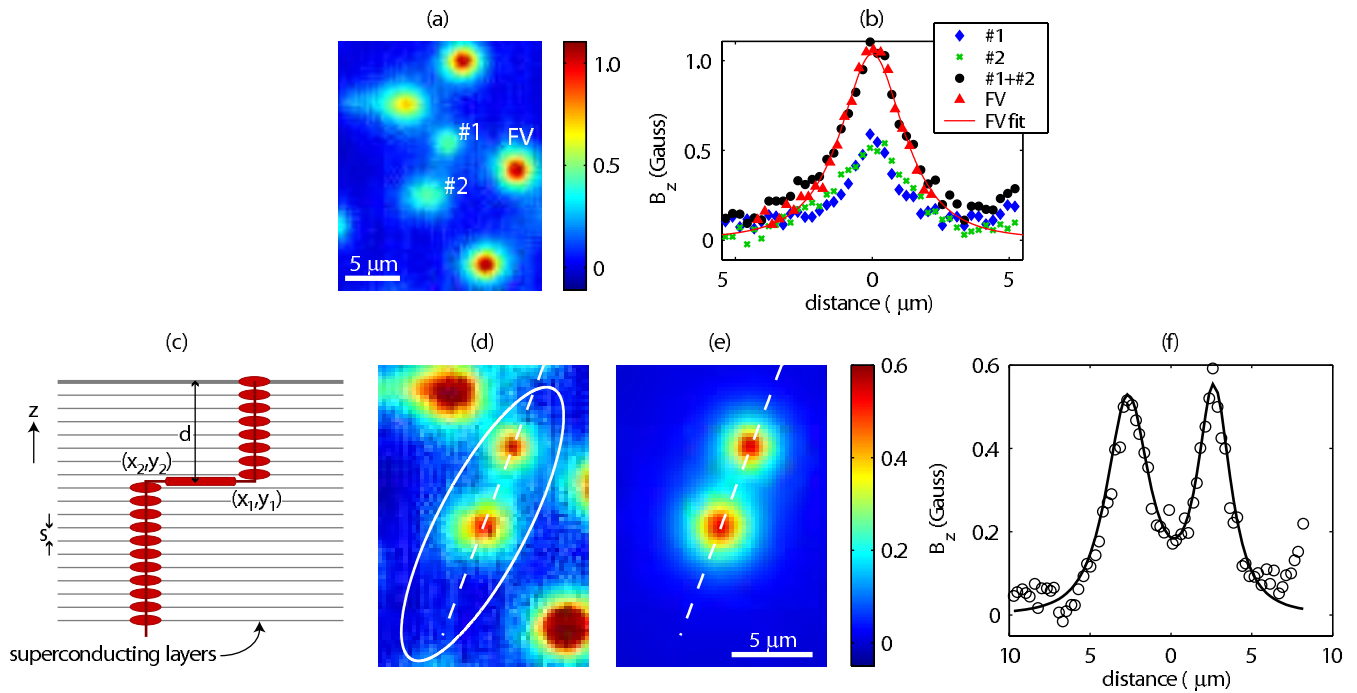}
\caption{\label{fig:PVfit}(Color).  Fit to a kinked pancake vortex
stack in the \ybco{6.375} at anneal stage 6 ($T_c=14.0$~K) at
$T=4$~K\@. (a) Hall probe image containing partial vortices
labeled \#1 and \#2. The 3 brightest features are full vortices.
(b) Horizontal cross sections through the partial vortices, the
sum of the partial vortices, and the full vortex labeled FV in
(a).  The solid line is from a 2D fit to the FV with
$z=1.0~\micron$ as a fixed parameter and
$\lambda_{ab}=0.65~\micron$ as a fit parameter.  (c) Sketch of the
assumed geometry of the kinked pancake stack, not to scale. (d)
Zoom in of image (a) with the color scale adapted to the signal
range of the partial vortices. (e) Fit to the partial vortices
within the white oval in (d) as discussed in the text. $d =
0.36~\micron$ and kink length is 5.3~\micron. (f) Cross-sections
through the data (circles) and fit (line) along the dashed lines
in (d) and (e).}
\end{figure*}

In Ref.\ \onlinecite{ClemJR:Two-di}, Clem introduced the idea of
2D pancake vortices as the basic building blocks of 3D vortices in
highly anisotropic superconductors consisting of weakly Josephson
coupled layers.  Even when the interlayer Josephson coupling is
not negligible, the vortex structure can be described as a
superposition of 2D pancake vortices and short sections of
Josephson vortices (called ``strings'') connecting pancakes in
adjacent layers.\cite{ClemJR:Panv} Clem\cite{ClemJR:Two-di} also
showed that for a vortex aligned along the $c$-axis, a straight
stack of 2D pancake vortices gives the same result as an ordinary
3D vortex in the anisotropic London model.  However, Benkraouda
and Clem\cite{BenkraoudaM:Instvl} proposed that a tilted pancake
stack may lower its energy by instead forming a kinked structure
similar to the one shown in Fig.\ \ref{fig:PVfit}(c), rather than
maintaining a homogeneous tilt angle.  Our observations of partial
vortices suggest such configurations of kinks and short pancake
stacks, which are stabilized by pinning effects.

Compared to optimally doped YBCO, where $\gamma =
\lambda_c/\lambda_{ab}\approx 5.5\pm 1$,\cite{Dolan_PRL_1989} our
highly underdoped \ybco{6+x} ($x\approx 0.35$--0.375) crystals are
much more anisotropic. Microwave measurements in similar samples
found zero temperature values $\lambda_c(0)\approx 100~\micron$
for $T_c\approx 6$~K, and $\lambda_c(0)\approx 40~\micron$ for
$T_c\approx 15$~K\@.\cite{HosseiniAR:lambdac} The zero temperature
in-plane penetration depth can be obtained from recent
measurements\cite{Liang_PRL_2005} of $H_{c1}(0)$ in similar
crystals which found that the power law $H_{c1}(0)=0.366\,
T_c^{1.64}$~(Oe) fit $H_{c1}$ vs.\ $T_c$ data well for $T_c\le
22$~K\@.  Using $H_{c1}=\Phi_0 (\mathrm{ln}(\kappa)+0.5)/(4\pi
\lambda_{ab}^2)$ and $\kappa \approx 40$ (Ref.\
\onlinecite{Gray_PRB_1992}) gives $\lambda_{ab}(0) =
1.00$~\micron{} and 0.47~\micron{} for $T_c=6$~K and 15~K,
respectively. These $\lambda_{ab}$ values are in agreement with
values we obtained from fits to Hall probe vortex images from our
variable $T_c$ \ybco{6.375} crystal as discussed in the Appendix.
Thus $\gamma \approx 100$ for $T_c \approx 6$~K and $\gamma
\approx 85$ for $T_c \approx 15$~K\@ in these crystals.

Since the 2D single layer screening length $\Lambda =
2\lambda_{ab}^2/s \approx 1$~mm is greater than $\lambda_c$ in our
highly underdoped YBCO crystals (the bilayer spacing in YBCO is
$s=1.17$~nm), interlayer Josephson coupling is not
negligible\cite{ClemJR:Two-di} and the notion of purely
magnetically coupled pancakes is not entirely accurate.  However,
due to the large values of $\lambda_c$ and $\gamma$, the
additional attraction between separated pancake stacks due to
Josephson strings is smaller than the magnetic interaction, as
will be discussed in Sec.\ \ref{sec:discussion}.  Thus the pancake
vortex plus Josephson string picture should be at least
qualitatively appropriate for these samples.

To show that short pancake stacks can explain our observations, we
consider the magnetic field that a straight partial stack extending from
$z_i$ to $z_f$ ($z_i < z_f \le 0$) generates above the $z=0$
surface of a layered superconductor assumed to be much thicker
than $\lambda_{ab}$. It was shown in Ref.\
\onlinecite{ClemJR:Panv} that the $z$ component of the magnetic
field at a height $z$ above the surface and a radius $r$ from the
vortex axis is
%%%%%%%%%%%%%%%%%%%%%%%%%%%%%%%%%%%%%%%%%%%
\begin{equation}
\label{eqn:shortstack-b}B_z(r,z) = \frac{\Phi_0}{2\pi
\lambda_{ab}^2} \int_0^\infty dq
    \frac{q\,e^{-qz}\,J_0(qr)}{Q(Q+q)}\left(e^{Qz_f} -
    e^{Qz_i}\right),
\end{equation}
%%%%%%%%%%%%%%%%%%%%%%%%%%%%%%%%%%%%%%%%%%%
where $Q = \sqrt{q^2+\lambda_{ab}^{-2}}$ and the layer spacing $s$
is much smaller than both $z$ and $\lambda_{ab}$.  If the partial
stack extends from $|z_i| \gg \lambda_{ab}$ to $z_f=0$, Eq.\
(\ref{eqn:shortstack-b}) gives the field of a conventional 3D
vortex (Eq.\ (\ref{eqn:fullmodel})). Integrating Eq.\
(\ref{eqn:shortstack-b}) gives total flux
%%%%%%%%%%%%%%%%%%%%%%%%%%%%%%%%%%%%%%%%%%%
\begin{equation}
\label{eqn:fluxshortstack}\Phi = \Phi_0\left(e^{z_f/\lambda_{ab}}
- e^{z_i/\lambda_{ab}}\right)
\end{equation}
%%%%%%%%%%%%%%%%%%%%%%%%%%%%%%%%%%%%%%%%%%%
through the $z=0$ surface of a superconducting half space for a
partial vortex extending from $z_i$ to
$z_f$.\cite{ClemJR:2Dpvfs,ClemJR:Panv} If an otherwise straight
vortex stack has one kink at a depth $z=-d$, the total flux
through the surface from the lower and upper partial stacks is
$\Phi_0 e^{-d/\lambda_{ab}}$ and $\Phi_0\left(1 -
e^{-d/\lambda_{ab}}\right)$, respectively. The total flux is
$\Phi_0$, as expected.

To compare the kinked stack model with our observations, we fit
the two partial vortices labeled in Fig.\ \ref{fig:PVfit}(a) to a
model of a pancake stack with one kink.
%%%%%%%%%%%%%%%%%%%%%%%%%%%%%%%%%%
%%%%%%%%%%%%%%%%%%%%%%%%%%%%%%%%%%
Figure \ref{fig:PVfit}(b) shows that partial vortex \#1 has a
narrower profile, so we chose it as the upper stack.  Adding the
field profiles (Eq.\ (\ref{eqn:shortstack-b})) of all the partial
stacks in a kinked stack gives the field profile of a straight
stack (full vortex). Figure \ref{fig:PVfit}(b) shows that the
cross section through a full vortex is similar to the sum of the
cross sections through partial vortices \#1 and \#2.

Using non-linear regression, we fit the portion of the image shown
in Fig.\ \ref{fig:PVfit}(d) inside the oval to a numerical
approximation of Eq.\ (\ref{eqn:shortstack-b}) with one expression
each for the lower ($-\infty$ to $-d$) and upper ($-d$ to 0)
partial pancake stacks.  Free fit parameters were the depth of the
split, $d$, and the radial centers ($x_1,y_1$) and ($x_2,y_2$) of
each partial stack (Fig.\ \ref{fig:PVfit}(c)). Fixed input values
were $z = 1.0~\micron$ and $\lambda_{ab} = 0.65~\micron$.  This $\lambda_{ab}$ was obtained
from a 2D fit to the full vortex labeled `FV' in Fig.\
\ref{fig:PVfit}(a) by the method described in the Appendix. Our
fit gives a depth of the kink $d = 0.36~\micron$ and lateral
displacement at the kink (kink length)
$\rho=\sqrt{(x_1-x_2)^2+(y_1-y_2)^2}=5.3~\micron$. Figure
\ref{fig:PVfit}(e) shows a 2D color plot of the fit and (f) shows
cross-sections of the data and fit images along the dotted lines.
Kinked or separated pancake vortex stacks is a plausible
interpretation of our observations since, as Fig.\ \ref{fig:PVfit}
shows, the model fits well to our data.

This partial vortex pair was ideal for fitting because there
appeared to be only a single kink within a few $\lambda_{ab}$ of
the surface and the partial stacks were well defined and circular.
In most of our partial vortex images there were multiple kinks in
a stack or several intermingled kinked stacks.  In these cases,
fitting the data would be more complicated. It was also common for
partial vortices to have non-uniform shapes (see Fig.\
\ref{fig:HPpartials}) which do not strictly agree with the model
of one or a few kinks in a pancake stack. Elongated partial
vortices and those with apparent tails, such as the partial vortex
to the upper left of \#1 in Fig.\ \ref{fig:PVfit}(a), could be the
result of many closely spaced kinks, a tilt of a partial stack, or
even a non-uniform staggering of pancakes from layer to layer.  In
each of these scenarios the displacement between adjacent pancakes
would be smaller than the Hall probe spatial resolution.  Though
the kinked stack model cannot be used to fit irregularly shaped
partial vortices without many free parameters, the underlying
phenomenon is similar, with 2D pancake vortices playing a critical
role.

\section{Other characteristics}\label{sec:furtherdetails}

A number of observed properties of these partial vortices further
substantiate the partial pancake stack interpretation as well as
give further insight into their stability and pinning. Partial
vortices occurred in groups, preferred certain regions in the
crystal, were more mobile than full vortices, and were more likely
to be formed by cooling in a tilted field.

The observed grouping of partial vortices is necessary for a
kinked pancake stack, since all pancake vortices near the crystal
surface carry flux through the surface which collectively adds to
$\Phi_0$.  We observed partial vortices up to tens-of-microns away
from others in a group, so even if a partial vortex appeared
isolated, other segments may have been outside the image area. One
caveat would be if there were subsurface vortex termination, which
could occur for small sample size.\cite{MintsRG:Vormcs}

Partial vortices showed a tendency to prefer certain regions of
the crystal, even after a room temperature annealing.  This was
especially noticeable for the later anneal stages (higher $T_c$)
for which partial vortices were rare.  This may indicate that
kinks occurred preferentially in regions of the sample which were
different from the bulk, perhaps with higher disorder, more
pinning sites, or weaker superconductivity.  At the lowest $T_c$
stage, we observed that partial vortices were more likely to pin
where the tip of the Hall probe sat when the probe's active area was centered over the scan area.  This was the location of the probe during cooldown through $T_c$ and also the place where the $z$ approach was most often performed to determine the location of the sample.  The vortices may have been preferentially attracted
to the location of the probe's tip, or the repeated contact with the sample in that location could have created
pinning sites.

\begin{figure}
\includegraphics{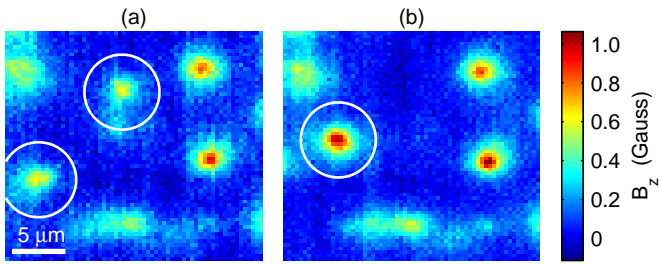}
\caption{\label{fig:twopartials2one}(Color). Partial vortices
coalesced during sample coarse motion ($T=2.4$~K)\@. (a) Hall
probe image after cooling through $T_c=8.6$~K in $H_z=0.21$~Oe
applied field. White circles identify two partial vortices. (b)
After slight $z$ coarse motion of the sample. Comparison of the
images suggests that the partial vortices in (a) collapsed to one
full vortex in (b).}
\end{figure}

We also found that partial vortices were more mobile than apparent
full vortices in the same samples.  For example, partial vortices
sometimes moved or coalesced after stick-slip coarse motion of the
sample holder in the $xy$ or $z$ directions. This was not observed
for full vortices. Figure \ref{fig:twopartials2one} shows Hall
probe images before and after several ramps of the voltage on the
$z$ piezoelectric and slight $z$ coarse motion.
%%%%%%%%%%%%%%%%%%%%%%%%%%%%%%%%%
%%%%%%%%%%%%%%%%%%%%%%%%%%%%%%%%%%
Two partial vortices of similar peak amplitude in Fig.\
\ref{fig:twopartials2one}(a) appeared to coalesce into one full
vortex in Fig.\ \ref{fig:twopartials2one}(b). Motion of partial
vortices during coarse motion could be due to stray fields from
the stick-slip high voltage pulses, which might create forces
large enough to unpin some partial vortices.  Once unpinned, a
kinked stack could realign to a straight stack as favored by
electromagnetic coupling of the pancakes and by any Josephson
coupling.

\begin{figure}
\includegraphics{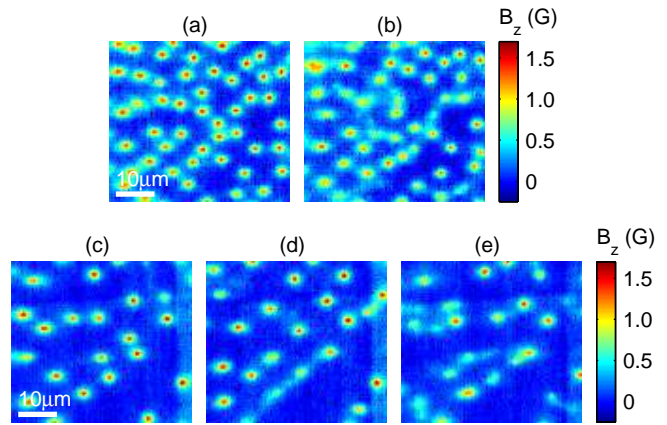}
\caption{\label{fig:Bxseries}(Color). Effect of an in-plane field
on partial vortex formation in \ybco{6.375}.  $T_c=14.4$~K and
$T=4$~K for all the images.  Field-cooled through $T_c$ in
$H_z=0.50$~Oe and (a) $H_x=0$, (b) $H_x=2.3$~Oe.  Field-cooled
through $T_c$ in $H_z=0.20$~Oe and (c) $H_x=0$, (d) $H_x=1.2$~Oe,
(e) $H_x=2.3$~Oe. Images acquired in zero applied field. The
$x$-direction is horizontal and $z$ is out of the page.}
\end{figure}

To encourage partial vortex formation, we cooled the sample in a
magnetic field with a horizontal component to reduce the energy
cost of a kink. With the \ybco{6.375} crystal almost fully
annealed with $T_c=14.4$~K (stage 7), the sample was cooled
through $T_c$ in an applied field $\vec{H} =
H_x\hat{x}+H_z\hat{z}$. The vortex arrangement did not change when
the field was turned off at 4~K\@.  The images in Fig.\
\ref{fig:Bxseries} show increased numbers of partial vortices with
increased $H_x$.
%%%%%%%%%%%%%%%%%%%%%%%%%%%%%%%%%%
%%%%%%%%%%%%%%%%%%%%%%%%%%%%%%%%%%
The magnitude of $H_z$ determined the density of flux observed in
the images. The horizontal field may have caused pancakes within a
stack to pin at large displacements with respect to each other as
$T$ was lowered through $T_c$. Since the flux arrangement did not
change when the field was turned off, the pinning must have been
sufficiently strong to overcome the restoring forces favoring a
straight stack.

\begin{figure}
\includegraphics{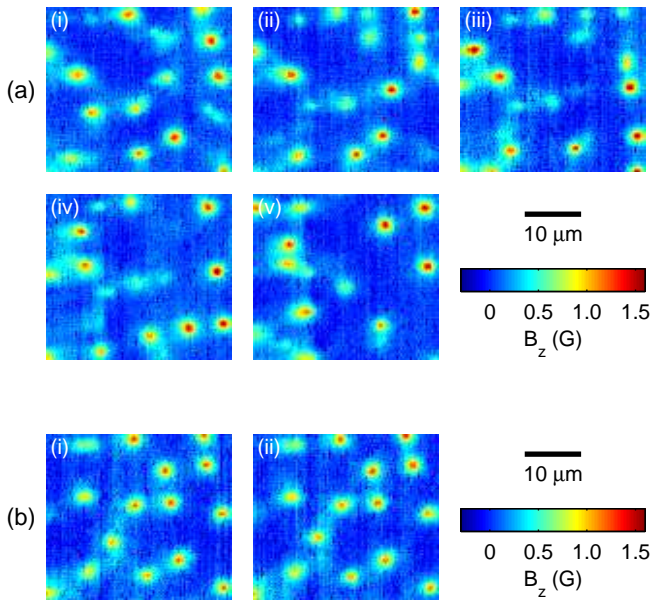}
\caption{\label{fig:thermalmotion}(Color). Comparison of vortex
movement after elevating the temperature for the \ybco{6.375}
crystal cooled in tilted and perpendicular fields.
$T_c=14.4$~K\@. (a) Field-cooled in $H_x=2.0$~Oe and $H_z=0.25$~Oe
to $T=3$~K\@. Image (i) was taken after the field was turned
off. After thermal cycling sequentially to (ii) 6.6~K, (iii)
8.0~K, (iv) 9.8~K, and (v) 12.2~K\@.  (b) The sample was
field-cooled in $H_x=0$ and $H_z=0.25$~Oe to $T=3$~K\@. Image (i)
was taken after the field was turned off. (ii) After thermal
cycling to 11.8~K\@. $T=3.0$~K for all images.}
\end{figure}

After cooling in a tilted field, we observed a change in the flux
arrangement after the temperature was raised but still kept below
$T_c$. As shown in Fig.\ \ref{fig:thermalmotion}(a), the
\ybco{6.375} sample with $T_c=14.4$~K (stage 7) was cooled to 3~K
in a field $\vec{H} = 2.0\hat{x}+0.25\hat{z}$~Oe, where $H_x$ was
chosen to induce partial vortices.  While at 3~K, the applied
field was turned off and no change was observed. Then the sample
was warmed to $T=6.6$~K, cooled back to 3~K, and imaged again.
This cycle was repeated several times with successively higher
maximum $T$. After all cycles, the flux arrangement changed.
%%%%%%%%%%%%%%%%%%%%%%%%%%%%%%%%%%
%%%%%%%%%%%%%%%%%%%%%%%%%%%%%%%%%%
Vortices created by cooling through $T_c$ in $\vec{H} =
0.25\hat{z}$~Oe did not show any motion after similar thermal
cycling (Fig.\ \ref{fig:thermalmotion}(b)).  These observations indicate high metastability of the vortices for cooling in non-zero $H_x$, as there are many possible structures consisting of pancake vortices and Josephson vortices in layered superconductors when cooled in a tilted field.

\section{Discussion}\label{sec:discussion}

The high-quality fit of the kinked stack model to our measured
field profiles (Sec.\ \ref{sec:themodel}), as well as the other
observed properties of the sub-$\Phi_0$ features (Sec.\
\ref{sec:furtherdetails}), suggest that we have observed kinked
pancake vortex stacks in the highly underdoped \ybco{6+x} crystals.
The large anisotropy and large $\lambda_c$ in these crystals
suggest that it is reasonable to think of vortices as being
composed of 2D pancake vortices.  Strictly speaking, however, the
Josephson coupling in these crystals is not negligible compared to
the magnetic coupling because $\lambda_c$ is \textsl{not} much
larger than $\Lambda$.\cite{ClemJR:Two-di}  The presence of
Josephson coupling will lead to some distortion of the field
profiles and to additional interaction energy, though a fully
formed Josephson vortex will not exist along the kink since the
kink length $\rho$ is typically less than $\lambda_c$. Despite
non-negligible Josephson coupling, the concept of pancake vortices
is still qualitatively valid here and, as we have shown in Sec.\
\ref{sec:themodel}, gives a good quantitative approximation of the
field profiles measured with the Hall probe.

Pinning is essential to explain our observations.  We know that
there are pinning sites in the YBCO crystals for all doping
values, as is typical of type II superconductors, because vortices
remain in the sample after the field is turned off below $T_c$. In
the absence of a field component in the $ab$-plane, magnetic
coupling of the pancake vortices and any interlayer Josephson
coupling favors alignment of the pancakes along the $c$-axis. That
we observe kinked pancake stacks in the highly underdoped YBCO even
after the applied field is turned off indicates that pinning of
the pancakes dominates the pancake vortex arrangement.  It should be noted that the calculations in Sec.\ \ref{sec:themodel} are for kinks in otherwise straight pancake vortex stacks, but pinning may cause fluctuations in the pancake positions within a stack, distorting the field profiles.\cite{GrigorenkoAN_PRB_2000}

To roughly quantify the strength of pinning required, we calculate
the restoring force on the kinked structure. From Ref.\
\onlinecite{BenkraoudaM:Instvl}, the energy required to deform a
straight stack to a singly-kinked structure (in the limit of zero
Josephson coupling and small layer spacing $s$) is
%%%%%%%%%%%%%%%%%%%%%%%%%%%%%%%%%%%%%%%%%%%
\begin{eqnarray}
E_\mathrm{kink}(\rho)&=&\left(\frac{\Phi_0}{2\pi}\right)^2
\frac{1}{4\lambda_{ab}} [(e^{-\rho/\lambda_{ab}}-1) +
\mathrm{ln}(\rho/\lambda_{ab})\nonumber\\
&&-~E_i(-\rho/\lambda_{ab}) + C],
\end{eqnarray}
%%%%%%%%%%%%%%%%%%%%%%%%%%%%%%%%%%%%%%%%%%%
where $E_i$ is the exponential-integral function and $C$ is
Euler's constant.  The restoring force on the kinked structure
with kink length $\rho$ found from $-dE_\mathrm{kink}/d\rho$ is
%%%%%%%%%%%%%%%%%%%%%%%%%%%%%%%%%%%%%%%%%%%
\begin{equation}
\label{eqn:Fkink}F_\mathrm{kink}(\rho) = -\left(\frac{\Phi_0}{4\pi
\lambda_{ab}}\right)^2 \left[\frac{\lambda_{ab}}{\rho} -
e^{-\rho/\lambda_{ab}}\left(1+\frac{\lambda_{ab}}{\rho}\right)
\right].
\end{equation}
%%%%%%%%%%%%%%%%%%%%%%%%%%%%%%%%%%%%%%%%%%%
For a kinked stack like that in Fig.\ \ref{fig:PVfit} with model
parameters $\rho=5.3~\micron$ and $\lambda_{ab}=0.65~\micron$,
this restoring force is 80~fN\@. This result is for a kink deep in
a crystal.

We can also consider the additional cost of a kinked stack due to
finite Josephson coupling.  In the limit $\gamma s < \rho <
\lambda_c$, which is the case for all the kinked stacks we
observed, Ref.\ \onlinecite{ClemJR:Panv} gives the energy cost of a long Josephson
string connecting the two partial stacks to be of the order $E_\mathrm{J} \approx
(\Phi_0/4\pi)^2 \rho (\lambda_{ab}\lambda_c)^{-1}$.  The
corresponding restoring force is of the order $F_\mathrm{J} \approx
-(\Phi_0/4\pi)^2 (\lambda_{ab}\lambda_c)^{-1}$. The $\lambda_c$
and $\lambda_{ab}$ values obtained from Refs.\
\onlinecite{HosseiniAR:lambdac} and \onlinecite{Liang_PRL_2005}
give this force to be approximately 3~fN for $T_c=6$~K and 14~fN
for $T_c=15$~K\@.  Comparing this to $F_\mathrm{kink}$ above, the
added restoring force due to the Josephson string is somewhat
smaller than that of the magnetic coupling.  For a longer Josephson string, the energy cost of the string increases linearly with the length and the restoring force $F_\mathrm{J}$ remains constant.  Thus, when the pinning force is large enough to compensate the restoring forces, there would be no theoretical limit to the maximum length of a metastable configuration.  As a point of
comparison with regards to pinning, measurements on a similar YBCO crystal with $T_c=11$~K
have estimated the required force to unpin a full vortex to be
$\sim$0.5~pN\@.\cite{GardnerBW:MansvY}

The calculated restoring forces would be much larger for a kinked stack in optimally doped YBCO\@.  Using the values of $\lambda_{ab} = 0.16$~\micron{} (Ref.\ \onlinecite{BasovDN:In-pla}) and $\gamma = 5.5$ (Ref.\ \onlinecite{Dolan_PRL_1989}) for near-optimally doped YBCO, $F_\mathrm{kink}$ has a maximum magnitude of 3~pN when $\rho = 2 \lambda_{ab}$, or if $\rho/\lambda_{ab} = 8.2$ as it was for the highly underdoped calculations above, $F_\mathrm{kink}$ would be 1 pN\@.  In the limit of a long Josephson string, the Josephson coupling gives an additional restoring force of magnitude $F_\mathrm{J} = 2$~pN\@.  These forces are much larger than for our highly underdoped YBCO\@, which is not surprising since optimally doped YBCO has much smaller anisotropy and much smaller $\lambda_c$, thus we expect the vortex behavior at higher doping to be less two-dimensional.  Indeed, we have never seen partial vortices in Ortho-II ($T_c\approx 60$~K) or near-optimally doped YBCO crystals.  Most of our measurements on these higher $T_c$ samples were after cooling in a perpendicular field.

Pinning in these highly underdoped crystals must be strong enough to
overcome the restoring force on the kinked pancake vortex stack.
The origin of the pinning sites cannot be determined from our Hall
probe images. They could be regions of oxygen inhomogeneity,
twinning boundaries, lattice imperfections, or something else.
Future studies of the pinning landscape in these crystals and in
other cuprates is desirable.  It is also not known how many of the
pancake vortices must be pinned to support the kinked structure in
the absence of an applied parallel field. However, the total force
due to pinning must be of order 80~fN to compensate for the
restoring force $F_\mathrm{kink}$. The data show that the pinning
becomes less sufficient for kinked vortex formation for the later
anneal stages (higher $T_c$).  As $T_c$ increases during the room
temperature annealing, $\lambda_{ab}$ decreases and so the restoring
force increases.  Also as the crystal anneals the oxygen chain
fragments get longer, and this may have an effect on the pinning.
However, since the crystals have Ortho-II ordering, there will
still be many oxygen vacancies even when fully annealed. The
number, spatial extent, and spatial distribution of vacancy
clusters may all be changing at once during an anneal. An open
question is whether oxygen inhomogeneities are necessary to see
partial vortices in these samples.

The behavior of the vortices when cooled in a magnetic field with
a parallel component (Figs.\ \ref{fig:Bxseries} and
\ref{fig:thermalmotion}) also bears discussion.  In a
perpendicular field in the absence of pinning, a vortex should
align parallel to the $c$-axis. When the field is tilted from the
perpendicular, the ground state for a vortex depends on the
parameters of the sample (see for example Ref.\
\onlinecite{Koshelev_PRB_2005:Vorcpl} for vortex lattices).  In
our YBCO sample $\gamma s < \lambda_{ab}$ and theory predicts a
transition from a tilted to a parallel vortex lattice as the field
approaches the $ab$-plane.\cite{Bulaevskii_PRB_1992} However,
Benkraouda and Clem showed that beyond small angles a kinked
structure can be energetically preferable to an isolated tilted
vortex, which itself is unstable beyond a tilt of
$52^{\circ}$.\cite{BenkraoudaM:Instvl} When our sample was cooled
through $T_c$ in a tilted field we saw many more partial vortices,
which we have suggested are kinked pancake vortex stacks. Since we
did not see any changes in our images when the field was turned
off at low temperature, pinning must have been sufficient to
sustain the pancake vortex arrangement. However, when the sample
was cycled to higher temperatures (still below $T_c$) the flux
arrangement changed as shown in Fig.\ \ref{fig:thermalmotion}(a).
We hypothesize that at the higher temperatures the pancake
vortices moved in an attempt to align along the $c$-axis, but in
some cases became pinned again as kinked stacks upon cooling. In
contrast, when the sample was cooled in a perpendicular field as
in Fig.\ \ref{fig:thermalmotion}(b), the vortices were already
pinned in their most stable state so no rearrangement occurred
when cycled to higher temperature.

The fact that the partial vortices formed even when the sample was
cooled in a perpendicular applied field along the $c$-axis (within
$\sim$1$^{\circ}$), especially for the lower $T_c$ values, further
indicates the importance of pinning. Many of the sub-$\Phi_0$
features are elongated or show tails, which could be due to
staggering of adjacent pancakes or many unresolvable kinks,
indicating a complex pinning landscape. Recent improvements in the
growth and preparation of these highly underdoped YBCO crystals have
resulted in better homogeneity and thus perhaps less pinning
compared to the earlier generation samples studied here. An ideal
sample to study would be a \ybco{6.333} crystal with Ortho-III
inverse ordering (every third chain full).  Such a highly underdoped
crystal would have very few chain vacancies.  Future work imaging
flux in the next generation samples is desirable.

A few other researchers have reported observations of kinked
pancake vortex stacks.  Grigorenko et
al.\cite{GrigorenkoAN:Visicv} reported a Hall probe image of one
``split'' pancake vortex stack in a
\mbox{Bi$_2$Sr$_2$CaCu$_2$O$_{8+\delta}$} (BSCCO) crystal with
$T_c=90$~K\@. The kinked stack in that case was formed under a
rapid change in magnetic field.  Extensive vortex imaging has been
done on
BSCCO\cite{Grigorenko_Nature_2001,GrigorenkoAN:Visicv,Bending_and_Dodgson_2005}
and typically kinked stacks such as ours have not been seen,
instead combined or crossing
lattices\cite{Bulaevskii_PRB_1992,Koshelev_PRL_1999} of pancake
vortex stacks and interlayer Josephson vortices are observed.
Unlike in BSCCO, we would not expect a crossing lattice to appear
as the applied field approaches the $ab$-plane in our highly underdoped YBCO crystals because the Josephson length $\gamma s$ is smaller than $\lambda_{ab}$.\cite{Bulaevskii_PRB_1992}

Beleggia et al.\cite{Beleggia} observed dumbbell-like features
consistent with kinked vortices in transmission electron
microscopy images of 300--400~nm thick films of optimally doped
YBCO when the applied field was within $7^{\circ}$ of parallel to
the film. Our work suggests that kinked vortices may form more
readily at low doping and can form even in the absence of an
applied parallel field.  The partial vortices we observed after cooling in only a perpendicular field could not have formed if sufficient pinning was not present.  However, it is not known if pinning is always required to form kinked stacks when a sample is in a continuously applied tilted field.

In conclusion, we have observed sub-$\Phi_0$ flux features in highly underdoped crystals of \ybco{6+x} with $T_c$ ranging from 5 to 15~K\@.  These ``partial vortices'' are well described as segments
of kinked stacks of 2D pancake vortices. The partial vortices were
more mobile than unkinked full vortices, formed more readily when
cooled in a magnetic field with a horizontal component, and were
seen most frequently for very low $T_c$.  Our observations provide
a view of vortex behavior in the highly underdoped region of the
YBCO phase diagram, showing that 2D vortex behavior and an
appropriate pinning landscape can produce complex flux features at
the crystal surface that are distinct from conventional vortices.

\begin{acknowledgments}
We thank J.R. Clem, J.R. Kirtley, S.A. Kivelson, and V.G. Kogan
for helpful discussions. Work at Stanford was funded by NSF Award
No.\ 9875193 and the DoE contract DE-AC02-76SF00515. Work at UBC was
funded by the CIAR and NSERC\@.
\end{acknowledgments}

\appendix*
% use the * to begin appendix section that has only 1 section.
\section{Penetration depth}

We also used Hall probe images of vortices in the variable $T_c$
\ybco{6.375} crystal to estimate the in-plane penetration depth
$\lambda_{ab}$, which relates to the superfluid density $n_s
\propto \lambda_{ab}^{-2}$, as a function of $T_c$ and $T$. Our
estimates deviate from the well established linearity between
$T_c$ and $n_s(0)$ first suggested by Uemura \textsl{et
al.}\cite{UemuraYJ:UnicbT,UemuraYJ:Bassac} for higher doped
cuprates.  Our results supplement $\lambda_{ab}$ values obtained
from $H_{c1}$ measurements in Ref.\ \onlinecite{Liang_PRL_2005}.
The unknown height of the Hall probe above the sample gives large
error bars on our $\lambda_{ab}$ results, and certain caveats
discussed below lead us to conservatively interpret our results as
upper bounds on $\lambda_{ab}$.

We chose a total of 40 different vortices in the \ybco{6.375}
sample for fitting, all of which appeared to be well-isolated full
vortices. Each vortex was fit with the anisotropic London model in
the thick crystal limit with the $ab$-plane parallel to the
surface:
\cite{PearlJ:Strsvn,KoganVG:Magfvc,KirtleyJR:Magfi-,KirtleyJR:Varsts}
%%%%%%%%%%%%%%%%%%%%%%%%%%%
\begin{equation}
\label{eqn:fullmodel}B_z(r,z) = \frac{\Phi_0}{2\pi
\lambda_{ab}^2}\int_0^\infty dq
\frac{q\,e^{-qz}\,J_0(qr)}{Q(Q+q)},
\end{equation}
%%%%%%%%%%%%%%%%%%%%%%%%%%%
where $Q = \sqrt{q^2+\lambda_{ab}^{-2}}$, $r$ is the radial
distance from the vortex axis, and $z$ is the height above the
sample surface.  We integrated Eq.\ (\ref{eqn:fullmodel}) at
constant $z$ over a $0.5~\mu$m diameter circular area representing
the Hall probe. The results are insensitive to the exact probe
size and shape.  We fit the vortex images using non-linear
regression to extract $\lambda_{ab}$ with fixed $z$. Free
parameters were the location of the vortex center, a constant
offset in the magnetic field (due to the Hall probe), and
$\lambda_{ab}$. The lengths $z$ and $\lambda_{ab}$ are strongly
correlated and could not both be free parameters.

In the scanning microscope $z=z_0+\Delta z$, where $z_0$ is the
sample-probe distance when touching, and $\Delta z$ is
controllable and for these measurements ranged from
0--0.16~\micron. Geometric constraints give a lower bound of
$z_0\geq 0.4~\micron$. A very conservative upper bound of $z_0\leq
1.45~\micron$ was obtained for this data set by fitting vortices
at maximum $T_c$ with $\lambda_{ab}=0$ and $z$ as a free
parameter. A smaller upper bound of $z_0\leq 1.3~\micron$ was
obtained by assuming $\lambda_{ab}(0)$ is at least as large as in
optimally doped YBCO\@.\cite{BasovDN:In-pla} All vortices were fit
with a range of $z_0$ values.  Fit results are reported here with
the typical value $z_0= 0.8~\micron$, with systematic error bars
determined by fits with $z_0= 0.4~\micron$ and $z_0= 1.3~\micron$.

\begin{figure}
\includegraphics{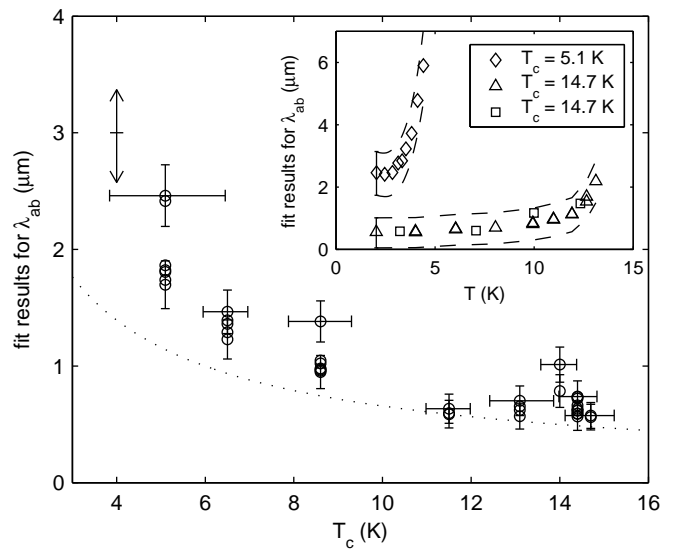}
\caption{\label{fig:lambdas}Fit results for low-temperature
$\lambda_{ab}$ versus $T_c$ for the \ybco{6.375} crystal, obtained
from fits to scanning Hall probe images of 40 different vortices
at eight $T_c$ stages of a room temperature annealing. The fitted
$\lambda_{ab}$ is an upper bound on the true $\lambda_{ab}(0)$.
Horizontal bars indicate the resolution-limited full width of the
superconducting transition for each $T_c$. Vertical error bars
(shown only for the extreme data points) are from uncertainty in
the probe calibration. A systematic error from uncertainty in the
minimum sample-probe distance $z_0$ could shift the full data set
by the extent indicated by the double arrow at $T_c=4$~K\@. Data
points shown are with $z_0 = 0.8~\micron$.  For comparison, the
dotted curve is from the fit to $H_{c1}(T_c)$ from Ref.\
\onlinecite{Liang_PRL_2005} as discussed in the text. Inset:
Temperature dependence of the $\lambda_{ab}$ results for two $T_c$
values. For each data set shown, $\lambda_{ab}$ was obtained from
fits to scanning Hall probe images of an individual vortex as $T$
increased.  The vortices disappeared before $T_c$.  The dashed
lines indicate the maximum extent the data sets could be shifted
due to systematic errors.}
\end{figure}

For four vortices we also took Hall probe images while warming to
investigate temperature dependence. The inset of Fig.\
\ref{fig:lambdas} shows $\lambda_{ab}(T)$ at the minimum and
maximum $T_c$ stages. Within our systematic and statistical
errors, it is not possible to extract the details of
$\lambda_{ab}(T)$ at low temperatures. The penetration depth
appears approximately constant for temperatures below $T_c/2$.
Thus we approximate $\lambda_{ab}(0)$ by
$\lambda_{ab}(T_\mathrm{min})$ with $T_\mathrm{min}\sim 2~K$ for
$T_c<12$~K and $T_\mathrm{min}\sim 4$~K for higher $T_c$.
%%%%%%%%%%%%%%%%%%%%%%%%%%%%%%%%%
%%%%%%%%%%%%%%%%%%%%%%%%%%%%%%%%%%

Our $\lambda_{ab}(T_\mathrm{min})$ versus $T_c$ results are shown
in Fig.\ \ref{fig:lambdas}. The dominant sources of error are the
uncertainty in $z_0$ (which will be the same for all data points)
and the calibration of the probe's Hall coefficient, $R_H=0.115\pm
0.015~\Omega$/G (which may fluctuate from cooldown to cooldown).
Fits with the $z_0$ extremes show that the corresponding
$\lambda_{ab}$ error, indicated by the double arrow in the upper
left of Fig.\ \ref{fig:lambdas}, is roughly constant for all data
points. The uncertainty in $R_H$ translates to a $\lambda_{ab}$
error of roughly $\pm 8\%$ of $z+\lambda_{ab}$. The in-plane
penetration depth decreased as $T_c$ increased.  For example, at
$T_c=6.5$~K, the average value is $\lambda_{ab}=1.35 \pm 0.18\
^{+0.37}_{-0.43}~\micron$, while for the highest $T_c = 14.7$~K it
is $\lambda_{ab} = 0.57 \pm 0.11\ ^{+0.34}_{-0.40} ~\micron$,
where the quoted errors are from the uncertainties in $R_H$ and
$z_0$, respectively.

%lambda_a and lambda_b anisotropy
The in-plane penetration depth $\lambda_{ab}$ is actually a
combination of $\lambda_a$ and $\lambda_b$, which are not equal in
YBCO\@.  The $\lambda_a/\lambda_b$ anisotropy has been measured to
be slightly larger than unity in higher doped samples (see for
example Ref.\ \onlinecite{Pereg-Barnea_PRB_2004}).  In-plane
anisotropy would lead to a distortion of the otherwise circular
field profile of a vortex aligned along the $c$-axis.  In
principle, our approach could be used to obtain separate values
for $\lambda_a$ and $\lambda_b$ if it was performed on a detwinned
crystal.

There are several caveats to our $\lambda_{ab}$ measurements.
First, it is possible that some of the full vortices identified
for fitting were actually partial vortices, especially at the
lowest $T_c$ values where fewer full vortices were available for
comparison. If a partial vortex stack was fit with Eq.\
(\ref{eqn:fullmodel}), the resulting $\lambda_{ab}$ value would be
falsely high.  At higher $T_c$, many more full vortices of
consistent appearance were observed. However, it cannot be ruled
out that the pancake vortices in the ``straight'' stacks were not
completely axial but rather pinned in a staggered manner resulting
in a more spread out vortex field profile.  If this were the case, our fits would yield falsely
high $\lambda_{ab}$ values since Eq.\ (\ref{eqn:fullmodel}) is for a straight vortex. Thus the most conservative approach to
our fit results is to take the values shown in Fig.\
\ref{fig:lambdas} as \textit{upper bounds} on the true values of
$\lambda_{ab}$.

For $T_c>10$~K, our $\lambda_{ab}(T_\mathrm{min})$ values obtained
from vortex fits are close to the values calculated from the low
$T_c$ fit $H_{c1}(0)=0.366\, T_c^{1.64}$~(Oe) from Ref.\
\onlinecite{Liang_PRL_2005}, as shown by the dotted curve in Fig.\
\ref{fig:lambdas}.  For $T_c<10$~K, our values lie slightly above
the curve.  However, as discussed above, our values should
conservatively be treated as upper bounds, so there is not a
discrepancy with Ref.\ \onlinecite{Liang_PRL_2005}.

Our $\lambda_{ab}$ estimates, along with the results of Ref.\
\onlinecite{Liang_PRL_2005}, give a larger $T_c$ for a given
superfluid density in the highly underdoped regime than predicted by
the Uemura relation $T_c \propto n_s(0)$.  Other recent
experiments have also shown deviations from this relation, such as
Pereg-Barnea \textsl{et al.}\cite{Pereg-Barnea_PRB_2004} in higher
doped YBCO crystals and Zuez \textsl{et
al.}\cite{Zuev_PRL_2005:Corbsd} in underdoped YBCO films. These
data indicate that thermal phase fluctuations alone cannot explain
the suppressed superfluid density in underdoped cuprates. Herbut
and Case\cite{Herbut_PRB_2004:Fintsd} proposed that low temperature nodal quasiparticles and
vortex fluctuations near $T_c$ can explain the observed
nonlinearity between $T_c$ and
$n_s(0)$.

\bibliography{GuikemaPRB_2008}% Produces the bibliography via BibTeX.

\begin{thebibliography}{38}
\expandafter\ifx\csname natexlab\endcsname\relax\def\natexlab#1{#1}\fi
\expandafter\ifx\csname bibnamefont\endcsname\relax
  \def\bibnamefont#1{#1}\fi
\expandafter\ifx\csname bibfnamefont\endcsname\relax
  \def\bibfnamefont#1{#1}\fi
\expandafter\ifx\csname citenamefont\endcsname\relax
  \def\citenamefont#1{#1}\fi
\expandafter\ifx\csname url\endcsname\relax
  \def\url#1{\texttt{#1}}\fi
\expandafter\ifx\csname urlprefix\endcsname\relax\def\urlprefix{URL }\fi
\providecommand{\bibinfo}[2]{#2}
\providecommand{\eprint}[2][]{\url{#2}}

\bibitem[{\citenamefont{Clem}(1991)}]{ClemJR:Two-di}
\bibinfo{author}{\bibfnamefont{J.~R.} \bibnamefont{Clem}},
  \bibinfo{journal}{Phys.\ Rev.\ B} \textbf{\bibinfo{volume}{43}},
  \bibinfo{pages}{7837} (\bibinfo{year}{1991}).

\bibitem[{\citenamefont{Grigorenko et~al.}(2002)\citenamefont{Grigorenko,
  Bending, Koshelev, Clem, Tamegai, and Ooi}}]{GrigorenkoAN:Visicv}
\bibinfo{author}{\bibfnamefont{A.~N.} \bibnamefont{Grigorenko}},
  \bibinfo{author}{\bibfnamefont{S.~J.} \bibnamefont{Bending}},
  \bibinfo{author}{\bibfnamefont{A.~E.} \bibnamefont{Koshelev}},
  \bibinfo{author}{\bibfnamefont{J.~R.} \bibnamefont{Clem}},
  \bibinfo{author}{\bibfnamefont{T.}~\bibnamefont{Tamegai}}, \bibnamefont{and}
  \bibinfo{author}{\bibfnamefont{S.}~\bibnamefont{Ooi}},
  \bibinfo{journal}{Phys.\ Rev.\ Lett.} \textbf{\bibinfo{volume}{89}},
  \bibinfo{pages}{217003} (\bibinfo{year}{2002}).

\bibitem[{\citenamefont{Beleggia et~al.}(2004)\citenamefont{Beleggia, Pozzi,
  Tonomura, Kasai, Matsuda, Harada, Akashi, Masui, and Tajima}}]{Beleggia}
\bibinfo{author}{\bibfnamefont{M.}~\bibnamefont{Beleggia}},
  \bibinfo{author}{\bibfnamefont{G.}~\bibnamefont{Pozzi}},
  \bibinfo{author}{\bibfnamefont{A.}~\bibnamefont{Tonomura}},
  \bibinfo{author}{\bibfnamefont{H.}~\bibnamefont{Kasai}},
  \bibinfo{author}{\bibfnamefont{T.}~\bibnamefont{Matsuda}},
  \bibinfo{author}{\bibfnamefont{K.}~\bibnamefont{Harada}},
  \bibinfo{author}{\bibfnamefont{T.}~\bibnamefont{Akashi}},
  \bibinfo{author}{\bibfnamefont{T.}~\bibnamefont{Masui}}, \bibnamefont{and}
  \bibinfo{author}{\bibfnamefont{S.}~\bibnamefont{Tajima}},
  \bibinfo{journal}{Phys.\ Rev.\ B} \textbf{\bibinfo{volume}{70}},
  \bibinfo{pages}{184518} (\bibinfo{year}{2004}).

\bibitem[{\citenamefont{Benkraouda and Clem}(1996)}]{BenkraoudaM:Instvl}
\bibinfo{author}{\bibfnamefont{M.}~\bibnamefont{Benkraouda}} \bibnamefont{and}
  \bibinfo{author}{\bibfnamefont{J.~R.} \bibnamefont{Clem}},
  \bibinfo{journal}{Phys.\ Rev.\ B} \textbf{\bibinfo{volume}{53}},
  \bibinfo{pages}{438} (\bibinfo{year}{1996}).

\bibitem[{\citenamefont{Geim et~al.}(2000)\citenamefont{Geim, Dubonos,
  Grigorieva, Novoselov, Peeters, and Schweigert}}]{Geim_Nature_2000}
\bibinfo{author}{\bibfnamefont{A.~K.} \bibnamefont{Geim}},
  \bibinfo{author}{\bibfnamefont{S.~V.} \bibnamefont{Dubonos}},
  \bibinfo{author}{\bibfnamefont{I.~V.} \bibnamefont{Grigorieva}},
  \bibinfo{author}{\bibfnamefont{K.~S.} \bibnamefont{Novoselov}},
  \bibinfo{author}{\bibfnamefont{F.~M.} \bibnamefont{Peeters}},
  \bibnamefont{and} \bibinfo{author}{\bibfnamefont{V.~A.}
  \bibnamefont{Schweigert}}, \bibinfo{journal}{Nature (London)}
  \textbf{\bibinfo{volume}{407}}, \bibinfo{pages}{55} (\bibinfo{year}{2000}).

\bibitem[{\citenamefont{Kirtley et~al.}(1995)\citenamefont{Kirtley, Chaudhari,
  Ketchen, Khare, Lin, and Shaw}}]{KirtleyJR:Dismfh}
\bibinfo{author}{\bibfnamefont{J.~R.} \bibnamefont{Kirtley}},
  \bibinfo{author}{\bibfnamefont{P.}~\bibnamefont{Chaudhari}},
  \bibinfo{author}{\bibfnamefont{M.~B.} \bibnamefont{Ketchen}},
  \bibinfo{author}{\bibfnamefont{N.}~\bibnamefont{Khare}},
  \bibinfo{author}{\bibfnamefont{S.-Y.} \bibnamefont{Lin}}, \bibnamefont{and}
  \bibinfo{author}{\bibfnamefont{T.}~\bibnamefont{Shaw}},
  \bibinfo{journal}{Phys.\ Rev.\ B} \textbf{\bibinfo{volume}{51}},
  \bibinfo{pages}{12057} (\bibinfo{year}{1995}).

\bibitem[{\citenamefont{Mannhart et~al.}(1996)\citenamefont{Mannhart,
  Hilgenkamp, Mayer, Gerber, Kirtley, Moler, and Sigrist}}]{MannhartJ:Genmfb}
\bibinfo{author}{\bibfnamefont{J.}~\bibnamefont{Mannhart}},
  \bibinfo{author}{\bibfnamefont{H.}~\bibnamefont{Hilgenkamp}},
  \bibinfo{author}{\bibfnamefont{B.}~\bibnamefont{Mayer}},
  \bibinfo{author}{\bibfnamefont{C.}~\bibnamefont{Gerber}},
  \bibinfo{author}{\bibfnamefont{J.~R.} \bibnamefont{Kirtley}},
  \bibinfo{author}{\bibfnamefont{K.~A.} \bibnamefont{Moler}}, \bibnamefont{and}
  \bibinfo{author}{\bibfnamefont{M.}~\bibnamefont{Sigrist}},
  \bibinfo{journal}{Phys.\ Rev.\ Lett.} \textbf{\bibinfo{volume}{77}},
  \bibinfo{pages}{2782} (\bibinfo{year}{1996}).

\bibitem[{\citenamefont{Liang et~al.}(1998)\citenamefont{Liang, Bonn, and
  Hardy}}]{LiangRuixing:GrohqY}
\bibinfo{author}{\bibfnamefont{R.}~\bibnamefont{Liang}},
  \bibinfo{author}{\bibfnamefont{D.~A.} \bibnamefont{Bonn}}, \bibnamefont{and}
  \bibinfo{author}{\bibfnamefont{W.~N.} \bibnamefont{Hardy}},
  \bibinfo{journal}{Physica (Amsterdam)} \textbf{\bibinfo{volume}{304C}},
  \bibinfo{pages}{105} (\bibinfo{year}{1998}).

\bibitem[{\citenamefont{Liang et~al.}(2002)\citenamefont{Liang, Bonn, Hardy,
  Wynn, Moler, Lu, Larochelle, Zhou, Greven, Lurio et~al.}}]{LiangR:PrechY}
\bibinfo{author}{\bibfnamefont{R.}~\bibnamefont{Liang}},
  \bibinfo{author}{\bibfnamefont{D.~A.} \bibnamefont{Bonn}},
  \bibinfo{author}{\bibfnamefont{W.~N.} \bibnamefont{Hardy}},
  \bibinfo{author}{\bibfnamefont{J.~C.} \bibnamefont{Wynn}},
  \bibinfo{author}{\bibfnamefont{K.~A.} \bibnamefont{Moler}},
  \bibinfo{author}{\bibfnamefont{L.}~\bibnamefont{Lu}},
  \bibinfo{author}{\bibfnamefont{S.}~\bibnamefont{Larochelle}},
  \bibinfo{author}{\bibfnamefont{L.}~\bibnamefont{Zhou}},
  \bibinfo{author}{\bibfnamefont{M.}~\bibnamefont{Greven}},
  \bibinfo{author}{\bibfnamefont{L.}~\bibnamefont{Lurio}},
  \bibnamefont{et~al.}, \bibinfo{journal}{Physica (Amsterdam)}
  \textbf{\bibinfo{volume}{383C}}, \bibinfo{pages}{1} (\bibinfo{year}{2002}).

\bibitem[{\citenamefont{Liang}()}]{Liang:PrivComm}
\bibinfo{author}{\bibfnamefont{R.}~\bibnamefont{Liang}}, \bibinfo{note}{private
  communication}.

\bibitem[{\citenamefont{Guikema}(2004)}]{GuikemaJW:THESIS}
\bibinfo{author}{\bibfnamefont{J.~W.} \bibnamefont{Guikema}},
  \bibinfo{howpublished}{Ph.D.\ Thesis, Stanford University}
  (\bibinfo{year}{2004}).

\bibitem[{\citenamefont{Chang et~al.}(1992)\citenamefont{Chang, Hallen,
  Harriott, Hess, Kao, Kwo, Miller, Wolfe, van~der Ziel, and
  Chang}}]{ChangAM:ScaHpm}
\bibinfo{author}{\bibfnamefont{A.~M.} \bibnamefont{Chang}},
  \bibinfo{author}{\bibfnamefont{H.~D.} \bibnamefont{Hallen}},
  \bibinfo{author}{\bibfnamefont{L.}~\bibnamefont{Harriott}},
  \bibinfo{author}{\bibfnamefont{H.~F.} \bibnamefont{Hess}},
  \bibinfo{author}{\bibfnamefont{H.~L.} \bibnamefont{Kao}},
  \bibinfo{author}{\bibfnamefont{J.}~\bibnamefont{Kwo}},
  \bibinfo{author}{\bibfnamefont{R.~E.} \bibnamefont{Miller}},
  \bibinfo{author}{\bibfnamefont{R.}~\bibnamefont{Wolfe}},
  \bibinfo{author}{\bibfnamefont{J.}~\bibnamefont{van~der Ziel}},
  \bibnamefont{and} \bibinfo{author}{\bibfnamefont{T.~Y.} \bibnamefont{Chang}},
  \bibinfo{journal}{Appl.\ Phys.\ Lett.} \textbf{\bibinfo{volume}{61}},
  \bibinfo{pages}{1974} (\bibinfo{year}{1992}).

\bibitem[{\citenamefont{Davidovi\'{c} et~al.}(1996)\citenamefont{Davidovi\'{c},
  Kumar, Reich, Siegel, Field, Tiberio, Hey, and Ploog}}]{DavidovicD:Cordam}
\bibinfo{author}{\bibfnamefont{D.}~\bibnamefont{Davidovi\'{c}}},
  \bibinfo{author}{\bibfnamefont{S.}~\bibnamefont{Kumar}},
  \bibinfo{author}{\bibfnamefont{D.~H.} \bibnamefont{Reich}},
  \bibinfo{author}{\bibfnamefont{J.}~\bibnamefont{Siegel}},
  \bibinfo{author}{\bibfnamefont{S.~B.} \bibnamefont{Field}},
  \bibinfo{author}{\bibfnamefont{R.~C.} \bibnamefont{Tiberio}},
  \bibinfo{author}{\bibfnamefont{R.}~\bibnamefont{Hey}}, \bibnamefont{and}
  \bibinfo{author}{\bibfnamefont{K.}~\bibnamefont{Ploog}},
  \bibinfo{journal}{Phys.\ Rev.\ Lett.} \textbf{\bibinfo{volume}{76}},
  \bibinfo{pages}{815} (\bibinfo{year}{1996}).

\bibitem[{\citenamefont{Oral et~al.}(1996)\citenamefont{Oral, Bending, and
  Henini}}]{OralA:ScaHpm}
\bibinfo{author}{\bibfnamefont{A.}~\bibnamefont{Oral}},
  \bibinfo{author}{\bibfnamefont{S.~J.} \bibnamefont{Bending}},
  \bibnamefont{and} \bibinfo{author}{\bibfnamefont{M.}~\bibnamefont{Henini}},
  \bibinfo{journal}{J. Vac.\ Sci.\ Technol.\ B} \textbf{\bibinfo{volume}{14}},
  \bibinfo{pages}{1202} (\bibinfo{year}{1996}).

\bibitem[{\citenamefont{Clem}(2004)}]{ClemJR:Panv}
\bibinfo{author}{\bibfnamefont{J.~R.} \bibnamefont{Clem}}, \bibinfo{journal}{J.
  Supercond.} \textbf{\bibinfo{volume}{17}}, \bibinfo{pages}{613}
  (\bibinfo{year}{2004}).

\bibitem[{\citenamefont{Dolan et~al.}(1989)\citenamefont{Dolan, Holtzberg,
  Feild, and Dinger}}]{Dolan_PRL_1989}
\bibinfo{author}{\bibfnamefont{G.~J.} \bibnamefont{Dolan}},
  \bibinfo{author}{\bibfnamefont{F.}~\bibnamefont{Holtzberg}},
  \bibinfo{author}{\bibfnamefont{C.}~\bibnamefont{Feild}}, \bibnamefont{and}
  \bibinfo{author}{\bibfnamefont{T.~R.} \bibnamefont{Dinger}},
  \bibinfo{journal}{Phys.\ Rev.\ Lett.} \textbf{\bibinfo{volume}{62}},
  \bibinfo{pages}{2184} (\bibinfo{year}{1989}).

\bibitem[{\citenamefont{Hosseini et~al.}(2004)\citenamefont{Hosseini, Broun,
  Sheehy, Davis, Franz, Hardy, Liang, and Bonn}}]{HosseiniAR:lambdac}
\bibinfo{author}{\bibfnamefont{A.}~\bibnamefont{Hosseini}},
  \bibinfo{author}{\bibfnamefont{D.~M.} \bibnamefont{Broun}},
  \bibinfo{author}{\bibfnamefont{D.~E.} \bibnamefont{Sheehy}},
  \bibinfo{author}{\bibfnamefont{T.~P.} \bibnamefont{Davis}},
  \bibinfo{author}{\bibfnamefont{M.}~\bibnamefont{Franz}},
  \bibinfo{author}{\bibfnamefont{W.~N.} \bibnamefont{Hardy}},
  \bibinfo{author}{\bibfnamefont{R.}~\bibnamefont{Liang}}, \bibnamefont{and}
  \bibinfo{author}{\bibfnamefont{D.~A.} \bibnamefont{Bonn}},
  \bibinfo{journal}{Phys.\ Rev.\ Lett.} \textbf{\bibinfo{volume}{93}},
  \bibinfo{pages}{107003} (\bibinfo{year}{2004}).

\bibitem[{\citenamefont{Liang et~al.}(2005)\citenamefont{Liang, Bonn, Hardy,
  and Broun}}]{Liang_PRL_2005}
\bibinfo{author}{\bibfnamefont{R.}~\bibnamefont{Liang}},
  \bibinfo{author}{\bibfnamefont{D.~A.} \bibnamefont{Bonn}},
  \bibinfo{author}{\bibfnamefont{W.~N.} \bibnamefont{Hardy}}, \bibnamefont{and}
  \bibinfo{author}{\bibfnamefont{D.}~\bibnamefont{Broun}},
  \bibinfo{journal}{Phys.\ Rev.\ Lett.} \textbf{\bibinfo{volume}{94}},
  \bibinfo{pages}{117001} (\bibinfo{year}{2005}).

\bibitem[{\citenamefont{Gray et~al.}(1992)\citenamefont{Gray, Kim, Veal,
  Seidler, Rosenbaum, and Farrell}}]{Gray_PRB_1992}
\bibinfo{author}{\bibfnamefont{K.~E.} \bibnamefont{Gray}},
  \bibinfo{author}{\bibfnamefont{D.~H.} \bibnamefont{Kim}},
  \bibinfo{author}{\bibfnamefont{B.~W.} \bibnamefont{Veal}},
  \bibinfo{author}{\bibfnamefont{G.~T.} \bibnamefont{Seidler}},
  \bibinfo{author}{\bibfnamefont{T.~F.} \bibnamefont{Rosenbaum}},
  \bibnamefont{and} \bibinfo{author}{\bibfnamefont{D.~E.}
  \bibnamefont{Farrell}}, \bibinfo{journal}{Phys.\ Rev.\ B}
  \textbf{\bibinfo{volume}{45}}, \bibinfo{pages}{10071} (\bibinfo{year}{1992}).

\bibitem[{\citenamefont{Clem}(1994)}]{ClemJR:2Dpvfs}
\bibinfo{author}{\bibfnamefont{J.~R.} \bibnamefont{Clem}},
  \bibinfo{journal}{Physica (Amsterdam)} \textbf{\bibinfo{volume}{235--240C}},
  \bibinfo{pages}{2607} (\bibinfo{year}{1994}).

\bibitem[{\citenamefont{Mints et~al.}(2000)\citenamefont{Mints, Kogan, and
  Clem}}]{MintsRG:Vormcs}
\bibinfo{author}{\bibfnamefont{R.~G.} \bibnamefont{Mints}},
  \bibinfo{author}{\bibfnamefont{V.~G.} \bibnamefont{Kogan}}, \bibnamefont{and}
  \bibinfo{author}{\bibfnamefont{J.~R.} \bibnamefont{Clem}},
  \bibinfo{journal}{Phys.\ Rev.\ B} \textbf{\bibinfo{volume}{61}},
  \bibinfo{pages}{1623} (\bibinfo{year}{2000}).

\bibitem[{\citenamefont{Grigorenko et~al.}(2000)\citenamefont{Grigorenko,
  Bending, Howells, and Humphreys}}]{GrigorenkoAN_PRB_2000}
\bibinfo{author}{\bibfnamefont{A.~N.} \bibnamefont{Grigorenko}},
  \bibinfo{author}{\bibfnamefont{S.~J.} \bibnamefont{Bending}},
  \bibinfo{author}{\bibfnamefont{G.~D.} \bibnamefont{Howells}},
  \bibnamefont{and} \bibinfo{author}{\bibfnamefont{R.~G.}
  \bibnamefont{Humphreys}}, \bibinfo{journal}{Phys.\ Rev.\ B}
  \textbf{\bibinfo{volume}{62}}, \bibinfo{pages}{721} (\bibinfo{year}{2000}).

\bibitem[{\citenamefont{Gardner et~al.}(2002)\citenamefont{Gardner, Wynn, Bonn,
  Liang, Hardy, Kirtley, Kogan, and Moler}}]{GardnerBW:MansvY}
\bibinfo{author}{\bibfnamefont{B.~W.} \bibnamefont{Gardner}},
  \bibinfo{author}{\bibfnamefont{J.~C.} \bibnamefont{Wynn}},
  \bibinfo{author}{\bibfnamefont{D.~A.} \bibnamefont{Bonn}},
  \bibinfo{author}{\bibfnamefont{R.}~\bibnamefont{Liang}},
  \bibinfo{author}{\bibfnamefont{W.~N.} \bibnamefont{Hardy}},
  \bibinfo{author}{\bibfnamefont{J.~R.} \bibnamefont{Kirtley}},
  \bibinfo{author}{\bibfnamefont{V.~G.} \bibnamefont{Kogan}}, \bibnamefont{and}
  \bibinfo{author}{\bibfnamefont{K.~A.} \bibnamefont{Moler}},
  \bibinfo{journal}{Appl.\ Phys.\ Lett.} \textbf{\bibinfo{volume}{80}},
  \bibinfo{pages}{1010} (\bibinfo{year}{2002}).

\bibitem[{\citenamefont{Basov et~al.}(1995)\citenamefont{Basov, Liang, Bonn,
  Hardy, Dabrowski, Quijada, Tanner, Rice, Ginsberg, and
  Timusk}}]{BasovDN:In-pla}
\bibinfo{author}{\bibfnamefont{D.~N.} \bibnamefont{Basov}},
  \bibinfo{author}{\bibfnamefont{R.}~\bibnamefont{Liang}},
  \bibinfo{author}{\bibfnamefont{D.~A.} \bibnamefont{Bonn}},
  \bibinfo{author}{\bibfnamefont{W.~N.} \bibnamefont{Hardy}},
  \bibinfo{author}{\bibfnamefont{B.}~\bibnamefont{Dabrowski}},
  \bibinfo{author}{\bibfnamefont{M.}~\bibnamefont{Quijada}},
  \bibinfo{author}{\bibfnamefont{D.~B.} \bibnamefont{Tanner}},
  \bibinfo{author}{\bibfnamefont{J.~P.} \bibnamefont{Rice}},
  \bibinfo{author}{\bibfnamefont{D.~M.} \bibnamefont{Ginsberg}},
  \bibnamefont{and} \bibinfo{author}{\bibfnamefont{T.}~\bibnamefont{Timusk}},
  \bibinfo{journal}{Phys.\ Rev.\ Lett.} \textbf{\bibinfo{volume}{74}},
  \bibinfo{pages}{598} (\bibinfo{year}{1995}).

\bibitem[{\citenamefont{Koshelev}(2005)}]{Koshelev_PRB_2005:Vorcpl}
\bibinfo{author}{\bibfnamefont{A.~E.} \bibnamefont{Koshelev}},
  \bibinfo{journal}{Phys.\ Rev.\ B} \textbf{\bibinfo{volume}{71}},
  \bibinfo{pages}{174507} (\bibinfo{year}{2005}).

\bibitem[{\citenamefont{Bulaevskii et~al.}(1992)\citenamefont{Bulaevskii,
  Ledvij, and Kogan}}]{Bulaevskii_PRB_1992}
\bibinfo{author}{\bibfnamefont{L.~N.} \bibnamefont{Bulaevskii}},
  \bibinfo{author}{\bibfnamefont{M.}~\bibnamefont{Ledvij}}, \bibnamefont{and}
  \bibinfo{author}{\bibfnamefont{V.~G.} \bibnamefont{Kogan}},
  \bibinfo{journal}{Phys.\ Rev.\ B} \textbf{\bibinfo{volume}{46}},
  \bibinfo{pages}{366} (\bibinfo{year}{1992}).

\bibitem[{\citenamefont{Grigorenko et~al.}(2001)\citenamefont{Grigorenko,
  Bending, Tamegai, Ooi, and Henini}}]{Grigorenko_Nature_2001}
\bibinfo{author}{\bibfnamefont{A.}~\bibnamefont{Grigorenko}},
  \bibinfo{author}{\bibfnamefont{S.}~\bibnamefont{Bending}},
  \bibinfo{author}{\bibfnamefont{T.}~\bibnamefont{Tamegai}},
  \bibinfo{author}{\bibfnamefont{S.}~\bibnamefont{Ooi}}, \bibnamefont{and}
  \bibinfo{author}{\bibfnamefont{M.}~\bibnamefont{Henini}},
  \bibinfo{journal}{Nature (London)} \textbf{\bibinfo{volume}{414}},
  \bibinfo{pages}{728} (\bibinfo{year}{2001}).

\bibitem[{\citenamefont{Bending and Dodgson}(2005)}]{Bending_and_Dodgson_2005}
\bibinfo{author}{\bibfnamefont{S.~J.} \bibnamefont{Bending}} \bibnamefont{and}
  \bibinfo{author}{\bibfnamefont{M.~J.~W.} \bibnamefont{Dodgson}},
  \bibinfo{journal}{J. Phys.: Condens. Matter} \textbf{\bibinfo{volume}{17}},
  \bibinfo{pages}{R955} (\bibinfo{year}{2005}).

\bibitem[{\citenamefont{Koshelev}(1999)}]{Koshelev_PRL_1999}
\bibinfo{author}{\bibfnamefont{A.~E.} \bibnamefont{Koshelev}},
  \bibinfo{journal}{Phys.\ Rev.\ Lett.} \textbf{\bibinfo{volume}{83}},
  \bibinfo{pages}{187} (\bibinfo{year}{1999}).

\bibitem[{\citenamefont{Uemura et~al.}(1989)\citenamefont{Uemura, Luke,
  Sternlieb, Brewer, Carolan, Hardy, Kadono, Kempton, Kiefl, Kreitzman
  et~al.}}]{UemuraYJ:UnicbT}
\bibinfo{author}{\bibfnamefont{Y.~J.} \bibnamefont{Uemura}},
  \bibinfo{author}{\bibfnamefont{G.~M.} \bibnamefont{Luke}},
  \bibinfo{author}{\bibfnamefont{B.~J.} \bibnamefont{Sternlieb}},
  \bibinfo{author}{\bibfnamefont{J.~H.} \bibnamefont{Brewer}},
  \bibinfo{author}{\bibfnamefont{J.~F.} \bibnamefont{Carolan}},
  \bibinfo{author}{\bibfnamefont{W.~N.} \bibnamefont{Hardy}},
  \bibinfo{author}{\bibfnamefont{R.}~\bibnamefont{Kadono}},
  \bibinfo{author}{\bibfnamefont{J.~R.} \bibnamefont{Kempton}},
  \bibinfo{author}{\bibfnamefont{R.~F.} \bibnamefont{Kiefl}},
  \bibinfo{author}{\bibfnamefont{S.~R.} \bibnamefont{Kreitzman}},
  \bibnamefont{et~al.}, \bibinfo{journal}{Phys.\ Rev.\ Lett.}
  \textbf{\bibinfo{volume}{62}}, \bibinfo{pages}{2317} (\bibinfo{year}{1989}).

\bibitem[{\citenamefont{Uemura et~al.}(1991)\citenamefont{Uemura, Le, Luke,
  Sternlieb, Wu, Brewer, Riseman, Seaman, Maple, Ishikawa
  et~al.}}]{UemuraYJ:Bassac}
\bibinfo{author}{\bibfnamefont{Y.~J.} \bibnamefont{Uemura}},
  \bibinfo{author}{\bibfnamefont{L.~P.} \bibnamefont{Le}},
  \bibinfo{author}{\bibfnamefont{G.~M.} \bibnamefont{Luke}},
  \bibinfo{author}{\bibfnamefont{B.~J.} \bibnamefont{Sternlieb}},
  \bibinfo{author}{\bibfnamefont{W.~D.} \bibnamefont{Wu}},
  \bibinfo{author}{\bibfnamefont{J.~H.} \bibnamefont{Brewer}},
  \bibinfo{author}{\bibfnamefont{T.~M.} \bibnamefont{Riseman}},
  \bibinfo{author}{\bibfnamefont{C.~L.} \bibnamefont{Seaman}},
  \bibinfo{author}{\bibfnamefont{M.~B.} \bibnamefont{Maple}},
  \bibinfo{author}{\bibfnamefont{M.}~\bibnamefont{Ishikawa}},
  \bibnamefont{et~al.}, \bibinfo{journal}{Phys.\ Rev.\ Lett.}
  \textbf{\bibinfo{volume}{66}}, \bibinfo{pages}{2665} (\bibinfo{year}{1991}).

\bibitem[{\citenamefont{Pearl}(1966)}]{PearlJ:Strsvn}
\bibinfo{author}{\bibfnamefont{J.}~\bibnamefont{Pearl}}, \bibinfo{journal}{J.
  Appl.\ Phys.} \textbf{\bibinfo{volume}{37}}, \bibinfo{pages}{4139}
  (\bibinfo{year}{1966}).

\bibitem[{\citenamefont{Kogan et~al.}(1993)\citenamefont{Kogan, Simonov, and
  Ledvij}}]{KoganVG:Magfvc}
\bibinfo{author}{\bibfnamefont{V.~G.} \bibnamefont{Kogan}},
  \bibinfo{author}{\bibfnamefont{A.~Y.} \bibnamefont{Simonov}},
  \bibnamefont{and} \bibinfo{author}{\bibfnamefont{M.}~\bibnamefont{Ledvij}},
  \bibinfo{journal}{Phys.\ Rev.\ B} \textbf{\bibinfo{volume}{48}},
  \bibinfo{pages}{392} (\bibinfo{year}{1993}).

\bibitem[{\citenamefont{Kirtley
  et~al.}(1999{\natexlab{a}})\citenamefont{Kirtley, Kogan, Clem, and
  Moler}}]{KirtleyJR:Magfi-}
\bibinfo{author}{\bibfnamefont{J.~R.} \bibnamefont{Kirtley}},
  \bibinfo{author}{\bibfnamefont{V.~G.} \bibnamefont{Kogan}},
  \bibinfo{author}{\bibfnamefont{J.~R.} \bibnamefont{Clem}}, \bibnamefont{and}
  \bibinfo{author}{\bibfnamefont{K.~A.} \bibnamefont{Moler}},
  \bibinfo{journal}{Phys.\ Rev.\ B} \textbf{\bibinfo{volume}{59}},
  \bibinfo{pages}{4343} (\bibinfo{year}{1999}{\natexlab{a}}).

\bibitem[{\citenamefont{Kirtley
  et~al.}(1999{\natexlab{b}})\citenamefont{Kirtley, Tsuei, Moler, Kogan, Clem,
  and Turberfield}}]{KirtleyJR:Varsts}
\bibinfo{author}{\bibfnamefont{J.~R.} \bibnamefont{Kirtley}},
  \bibinfo{author}{\bibfnamefont{C.~C.} \bibnamefont{Tsuei}},
  \bibinfo{author}{\bibfnamefont{K.~A.} \bibnamefont{Moler}},
  \bibinfo{author}{\bibfnamefont{V.~G.} \bibnamefont{Kogan}},
  \bibinfo{author}{\bibfnamefont{J.~R.} \bibnamefont{Clem}}, \bibnamefont{and}
  \bibinfo{author}{\bibfnamefont{A.~J.} \bibnamefont{Turberfield}},
  \bibinfo{journal}{Appl.\ Phys.\ Lett.} \textbf{\bibinfo{volume}{74}},
  \bibinfo{pages}{4011} (\bibinfo{year}{1999}{\natexlab{b}}).

\bibitem[{\citenamefont{Pereg-Barnea et~al.}(2004)\citenamefont{Pereg-Barnea,
  Turner, Harris, Mullins, Bobowski, Raudsepp, Liang, Bonn, and
  Hardy}}]{Pereg-Barnea_PRB_2004}
\bibinfo{author}{\bibfnamefont{T.}~\bibnamefont{Pereg-Barnea}},
  \bibinfo{author}{\bibfnamefont{P.~J.} \bibnamefont{Turner}},
  \bibinfo{author}{\bibfnamefont{R.}~\bibnamefont{Harris}},
  \bibinfo{author}{\bibfnamefont{G.~K.} \bibnamefont{Mullins}},
  \bibinfo{author}{\bibfnamefont{J.~S.} \bibnamefont{Bobowski}},
  \bibinfo{author}{\bibfnamefont{M.}~\bibnamefont{Raudsepp}},
  \bibinfo{author}{\bibfnamefont{R.}~\bibnamefont{Liang}},
  \bibinfo{author}{\bibfnamefont{D.~A.} \bibnamefont{Bonn}}, \bibnamefont{and}
  \bibinfo{author}{\bibfnamefont{W.~N.} \bibnamefont{Hardy}},
  \bibinfo{journal}{Phys.\ Rev.\ B} \textbf{\bibinfo{volume}{69}},
  \bibinfo{pages}{184513} (\bibinfo{year}{2004}).

\bibitem[{\citenamefont{Zuev et~al.}(2005)\citenamefont{Zuev, Kim, and
  Lemberger}}]{Zuev_PRL_2005:Corbsd}
\bibinfo{author}{\bibfnamefont{Y.}~\bibnamefont{Zuev}},
  \bibinfo{author}{\bibfnamefont{M.~S.} \bibnamefont{Kim}}, \bibnamefont{and}
  \bibinfo{author}{\bibfnamefont{T.~R.} \bibnamefont{Lemberger}},
  \bibinfo{journal}{Phys.\ Rev.\ Lett.} \textbf{\bibinfo{volume}{95}},
  \bibinfo{pages}{137002} (\bibinfo{year}{2005}).

\bibitem[{\citenamefont{Herbut and Case}(2004)}]{Herbut_PRB_2004:Fintsd}
\bibinfo{author}{\bibfnamefont{I.~F.} \bibnamefont{Herbut}} \bibnamefont{and}
  \bibinfo{author}{\bibfnamefont{M.~J.} \bibnamefont{Case}},
  \bibinfo{journal}{Phys.\ Rev.\ B} \textbf{\bibinfo{volume}{70}},
  \bibinfo{pages}{094516} (\bibinfo{year}{2004}).

\end{thebibliography}

\end{document}